\documentclass[12pt]{article}

\title{Quantum Calculus of Fibonacci Divisors and Fermion-Boson Entanglement for Infinite Hierarchy of N = 2 Supersymmetric Golden Oscillators}

\author{Oktay K Pashaev \\
Department of Mathematics \\ Izmir Institute of Technology \\ Izmir 35430, Turkey}

\begin{document}

\maketitle              % typeset the title of the contribution

\begin{abstract}
  The quantum calculus with two bases, as powers of the Golden and the Silver  ratio, relates Fibonacci divisor
	derivative with Binet formula of Fibonacci divisor number operator, acting in Fock space of quantum states.
It provides a tool to study the hierarchy of Golden oscillators with energy spectrum in form of Fibonacci divisor numbers. We generalize this model to supersymmetric number operator and corresponding Binet formula for supersymmetric Fibonacci divisor number operator. The operator determines the Hamiltonian of hierarchy of supersymmetric Golden oscillators, acting in 
fermion-boson Hilbert space and belonging to N=2 supersymmetric algebra.  The eigenstates of the super Fibonacci divisor number operator are double degenerate and can be characterized by a point on the super-Bloch sphere. By the supersymmetric Fibonacci divisor annihilation operator, we construct the hierarchy of supersymmetric coherent states as eigenstates of this operator. Entanglement of fermions with bosons in these states is calculated by the concurrence, represented by the Gram determinant and hierarchy of Golden exponential functions.  
We show that the reference states and corresponding von Neumann entropy, measuring fermion-boson entanglement,  are characterized completely by  the powers of the Golden ratio. The simple geometrical 
classification of entangled states by the Frobenius ball and
meaning of the concurrence as double area of parallelogram in Hilbert space are given.
\end{abstract}


%\footnote{An example footnote.}

 %
\section{Introduction}
The quantum calculus, as calculus of finite differences under scaling transformation, becomes important tool to study representation of quantum groups, arising in quantum integrable systems.
Historically, Pythagoras string and Fermat partition preceded classical works of Euler, Gauss, Jackson, Jacobi, Ramanujan etc. in this calculus. After discovering quantum integrable models it has attracted
many researchers, looking for $q$-deformations of physical models. 
The key point of this calculus is the $q$-number $[n]_q$, as deformation of natural numbers by geometric progression and corresponding $q$-derivative $D_q$.  The last one is derived as function of
dilatation operator $x \frac{d}{d x}$, so that the operator $q$-number $x D_q = [x \frac{d}{d x}]_q$ represents $q$-deformed dilatation operator. 

The dilatation operator plays central role also in recent attempts to formulate quantum mechanical 
solution of the Riemann zeta function problem, where
 nontrivial zeros of Riemann zeta function can be interpreted as spectrum of energy levels of some Hermitian operator in Hilbert space (Polya, Hilbert).
In \cite{Berry}, the scaling or dilatation operator for the classical Hamiltonian $H_{BK} = q p$ ("the Riemann dynamics")
was proposed. 
It turns out that this classical Hamiltonian $H_{BK}$ can be related with some physical systems.
In \cite{Blazone}, the damped harmonic oscillator was studied in Lagrangian formalism with two time-reversal equations with positive and negative damping. 
The equations of motion are modeling several physical systems, as topologically massive Chern-Simons gauge theory in pseudo-Euclidean space and dynamics of Bloch
electron, when the Fermi surface is near the band boundary, and the sign of one of the mass tensor component is negative.  In the limit of strong damping, which corresponds to the so-called Chern-Simons massless limit, valid for long wave excitations, the degenerate system has Hamiltonian $H =k\, x y$ of the Riemann dynamics, 
with Poisson bracket $\{x, y \} = -\frac{1}{\gamma}$, satisfying 
first order equations $\dot{x} = - \frac{k}{\gamma} x$, $\dot{y} = \frac{k}{\gamma} y$, with solution $x(t) = x(0) \exp (-\frac{k}{\gamma} t)$, $y(t) = y(0) \exp (\frac{k}{\gamma} t)$ \cite{Blazone}.

The quantum mechanical version of $H_{BK}$ in Schr\"odinger representation, takes form of dilatation operator, so that the Hamiltonian operator $h_{BK} = \frac{1}{2} (x p + p x) = -i\hbar (x \frac{d}{dx} +\frac{1}{2})$ ("the Riemann operator") 
and it was studied in \cite{Berry}.
In \cite{Bender} the operator valued zeta function $\zeta(\frac{1}{2}(1 - i h_{BK}) )$ was introduced and  this specific form of operator function can be interpreted as $f(s) = \zeta(s)$ deformation of original operator
$\frac{1}{2}(1 - i h_{BK})$, corresponding to the $f$ deformation of classical damped harmonic oscillator in strong damping limit. The generic $f$ deformation of harmonic oscillator \cite{Manko1}, \cite{Manko2} describes nonlinear oscillator with frequency, depending on amplitude, adiabatic invariant or integral of motion \cite{P3}, and it  includes in particular $q$-deformed and $pq$-deformed oscillators. Recently, a supersymmetric quantum
mechanical model based on function of dilatation operator, in which the eigenvalues of the
Hamiltonian are given in terms of Riemann zeta
functions was proposed in \cite{K}. Due to supersymmetry, the
ground state energy in the model is zero, which, as shown leads
to zeros of the zeta function on critical line.

It is interesting to note that the complex analytic form of dilatation operator 
$z \frac{d}{dz}$ and its q-deformation $z D_q = [z \frac{d}{dz}]_q$, where $z=x+i y$, acting on an analytic function $f(z)$, has appeared in planar hydrodynamic problem
of incompressible and irrotational flow in annular domain \cite{P8}. According to  the two circle theorem \cite{P2}, the flow is determined by an infinite set of images, 
distributed as Fermat partition for successive inversions in two circles, described by q-periodic 
functions.

In quantum theory of harmonic oscillator, this dilatation operator 
$z \frac{d}{dz}$ 
is just the number operator $N$  in Fock-Bargman representation, generating dilatation in complex plane $\alpha$, for coherent state $|\alpha\rangle$, so that $q^{N} |\alpha\rangle = |q\alpha\rangle$.
Then, the $pq$-deformation of number operator is given by 
$pq$-derivative operator determined by  $[N]_{pq} = z D_{pq}$.  
Corresponding  quantum calculus with two bases numbers, $p$ and $q$, called also as the "post-quantum" or  $pq$-quantum calculus \cite{CJ}, \cite{A}, admits several reductions for specific values of 
the numbers. For $p = \varphi$, $q=\varphi'$, where $\varphi = \frac{1+\sqrt{5}}{2}$, 
$\varphi' = \frac{1-\sqrt{5}}{2} = - \varphi^{-1}$ are the Golden and the Silver ratio correspondingly, this calculus was coined as the Binet-Fibonnaci calculus or shortly, 
 the Golden calculus \cite{P1}.  
In this case the Binet formula for Fibonacci numbers 
\begin{equation}
F_n = \frac{\varphi^n - {\varphi'}^n}{\varphi - \varphi'} = [n]_{\varphi \varphi'}
\end{equation}
 takes the form of $\varphi \varphi'$
quantum calculus number $[n]_{\varphi \varphi'}$. The formula was extended to Fibonacci number operator $F_N = [N]_{\varphi \varphi'}$, acting in the
Fock-Bargmann representation as $z D_F = F_{z \frac{d}{dz}} = [z \frac{d}{dz}]_{\varphi \varphi'}$, where
the Golden derivative $D_F$, becomes the central object 
of Golden calculus \cite{P1}. The Hamiltonian of quantum Golden oscillator is defined by $F_N$ and the corresponding spectrum is determined by Fibonacci numbers. 
It is known that Fibonacci numbers with multiplicative index $F_{kn}$ are dividable by 
$F_k$, and the division gives the hierarchy of sequences of integer numbers   $F_n^{(k)} = F_{n k}/F_k$, called as the higher-Fibonacci numbers \cite{P1} or the Fibonacci divisors \cite{P4}.
The first member of the hierarchy with $k=0$ is the set of integer numbers $n = \lim_{k\rightarrow 0} F^{(k)}_n$. For $k=1$ it corresponds to Fibonacci sequence $F_n = F^{(1)}_n$, and for every $k > 1$,  we get the sequences, satisfying
generalized Fibonacci triple relations.  The quantum calculus of Fibonacci divisors and the corresponding quantum mechanical Hamiltonian were studied in \cite{P4}.  In \cite{P7} Fibonacci numbers and their generalizations were derived in quantum coin flipping problem, and  quantum calculus of Fibonacci divisors was related with method of images in planar hydrodynamics in \cite{P9}.

Very recently \cite{P6}, the Golden calculus and Golden oscillator \cite{P1} were generalized to the supersymmetric case, 
by using supersymmetric number operator ${\cal N}$ and corresponding Binet formula for supersymmetric Fibonacci operator ${\cal F}_{\cal N}$. 

In present paper we generalize construction \cite{P6} to the hierarchy of supersymmetric Golden oscillators, determined by Fibonacci divisors. 
By the Binet formula we define the supersymmetric operator for Fibonacci divisors ${\cal F}^{(k)}_{\cal N}$ and corresponding Hamiltonian.  
The Hamiltonian  is acting in ${\cal H}_f \otimes {\cal H}_b$ - fermion-boson Hilbert space and belongs to $N=2$ supersymmetric algebra. In particular cases 
we have reductions to supersymmetric oscillator ($k=0$) and to supersymmetric Golden oscillator ($k=1$), \cite{P6}. 
Moreover, by taking partial trace according to fermionic degrees of freedom our Hamiltonian reduces to the hierarchy of Golden oscillators \cite{P4}. It turns out that the eigenstates of the super Fibonacci divisor number operator are double degenerate and can be characterized by points of the super-Bloch sphere. By introducing the supersymmetric Fibonacci divisor annihilation operator, 
the hierarchy of supersymmetric Golden coherent states $|\beta_k\rangle$ is obtained as eigenstates of this operator. 
Depending on position on the super-Bloch sphere, these states can be entangled. 
We calculate the fermion-boson entanglement in these states by using the concurrence, as the linear entropy, represented by the Gram determinant of the inner products, determined  by hierarchy of the Golden exponential functions. In \cite{P4} these functions were introduced for description of inner product of hierarchy of Golden coherent states in the Fock-Bargmann representation. 
 The supersymmetric coherent states $|\beta_k\rangle$ in the limit $\beta_k \rightarrow 0$ reduce to the reference states, entanglement of which is calculated by the von Neumann entropy, measuring fermion-boson entanglement. It is shown that the entanglement is characterized completely by the Golden ratio only. We give also simple geometrical 
classification of entangled states by the Frobenius ball and
meaning of the concurrence as double area of parallelogram in Hilbert space.
 
The paper is organized as follows. In Section 2 we briefly review main properties of Fibonacci divisor hierarchy of Golden oscillators. The hierarchy of supersymmetric Golden oscillator is studied in
Section 3. By introducing supersymmetric annihilation operator, in Section 4 we describe the hierarchy of Golden coherent states. Entanglement and entropy of these states and corresponding reference states
are studied in Section 5. 

\section{Hierarchy of Golden Quantum Oscillators}

  Here, we briefly recall basic properties of  the hierarchy of Golden oscillators, introduced in  \cite{P4}. It is determined by the pair of creation and annihilation operators, $b^\dagger_k$ and $b_k$, and number operator $N$, satisfying the quantum algebra
\begin{eqnarray}
b_k b^\dagger_k - \varphi^k b^\dagger_k b_k = \varphi'^{k N}, \hskip1cm
b_k b^\dagger_k - {\varphi'}^k b^\dagger_k b_k = \varphi^{k N},  
\end{eqnarray}
and  $[N,b^\dagger_k] = b^\dagger_k$,     $[N, b_k] = -b_k$. 
In the limit $k \rightarrow 0$, this algebra becomes just the bosonic algebra, $[b, b^\dagger] =I$ for $b \equiv b_0$, $b^\dagger \equiv b_0^\dagger$. For $k =1$ it takes form of the quantum algebra 
for the Golden oscillator \cite{P1}.
The Fibonacci divisor number operator is determined by operator version of the Binet type formula
\begin{equation}
F^{(k)}_N = \frac{(\varphi^k)^N - ({\varphi'}^k)^N}{\varphi^k - {\varphi'}^k} = [N]_{\varphi^k {\varphi'}^k}\,,
\end{equation}
and it has the meaning of the operator $pq$-number with bases, as the k-th power of Golden and Silver ratio $\varphi^k$ and ${\varphi'}^k$, correspondingly.
It can be represented by Fibonacci number operator 
\begin{equation}
F^{(k)}_N = \frac{1}{F_k} F_{k N}\,,\label{Fkn}
\end{equation}
and in factorized form as
\begin{eqnarray}
F^{(k)}_N = b^\dagger_k b_k\,, \hskip1cm F^{(k)}_{N+I} = b_k b^\dagger_k\,.
\end{eqnarray}
The operator satisfies recursion relations
\begin{eqnarray}
F^{(k)}_{N+I} &=& L_k F^{(k)}_N + (-1)^{k-1} F^{(k)}_{N-I}, \\
(\varphi^k)^N &=& \varphi^k F^{(k)}_N + (-1)^{k+1}F^{(k)}_{N-I}, \\
({\varphi'}^k)^N &=& {\varphi'}^k F^{(k)}_N + (-1)^{k+1}F^{(k)}_{N-I},\\
F^{(k)}_{N+I} &=& \varphi^k F^{(k)}_N + {\varphi'}^{k N}, \\
F^{(k)}_{N+I} &=& {\varphi'}^k F^{(k)}_N + {\varphi}^{k N},
\end{eqnarray}
gives commutator
\begin{eqnarray}
    b_k b^\dagger_k - b^\dagger_k b_k = [b_k, b^\dagger_k] = F^{(k)}_{N+I} - F^{(k)}_{N},
\end{eqnarray}
and
\begin{eqnarray}
[F^{(k)}_N, {b_k^\dagger}_F] &=& (F^{(k)}_N - F^{(k)}_{N-I}) {b_k^\dagger} = {b_k^\dagger} (F^{(k)}_{N+I} - F^{(k)}_{N}) , \\
b_k^\dagger f(F^{(k)}_{N+I}) &=& f(F^{(k)}_N) b_k^\dagger, \\
b_k f(F^{(k)}_{N}) &=& f(F^{(k)}_{N+I}) b_k.
\end{eqnarray}
 The Fock space basis states  $ {\cal H} =\{ |n; k\rangle   \}$, defined as
\begin{equation}
|n; k\rangle = \frac{(b^\dagger_k)^n}{\sqrt{F^{(k)}_n!}} |0; k\rangle,\hskip1cm b_k |0; k\rangle =0,\label{deformedFock}
\end{equation}
are orthonormal $\langle n; k | m; k \rangle = \delta_{n m}$ eigenstates of 
$F^{(k)}_N$,
\begin{eqnarray}
F^{(k)}_N |n;k\rangle = F^{(k)}_n |n;k\rangle\,, 
\end{eqnarray}
 and satisfy relations
\begin{equation}
   b^\dagger_k |n; k\rangle = \sqrt{F^{(k)}_{n+1}} |n+1; k\rangle, \hskip0.5cm b_k |n; k\rangle = \sqrt{F^{(k)}_{n}}|n-1; k\rangle\,.
\end{equation}
The creation/annihilation operators $b^\dagger_k$, $b_k$ can be connected with bosonic operators $b^\dagger$, $b$ by nonlinear transformation
\begin{equation}
b_k = b \sqrt{\frac{F^{(k)}_N}{N}} = \sqrt{\frac{F^{(k)}_{N+I}}{N+I}} b, \hskip1cm b^\dagger_k = \sqrt{\frac{F^{(k)}_N}{N}} b^\dagger = b^\dagger\sqrt{\frac{F^{(k)}_{N+I}}{N+I}},
\end{equation}
or due to (\ref{Fkn})
\begin{equation}
b_k = b \sqrt{\frac{F_{k N}}{F_k N}} = \sqrt{\frac{F_{k (N+I)}}{F_k (N+I)}} b, \hskip1cm b^\dagger_k = \sqrt{\frac{F_{k N}}{F_k N}} b^\dagger = b^\dagger\sqrt{\frac{F_{k(N+I)}}{F_k (N+I)}}.
\end{equation}

The spectrum of the hierarchy of Golden deformed bosonic Hamiltonians, defined as
\begin{equation}
H_k = \frac{\hbar \omega}{2} (b_k b^\dagger_k + b^\dagger_k b_k) = \frac{\hbar \omega}{2} (F^{(k)}_N + F^{(k)}_{N+I}) \label{goldenH}
\end{equation}
is not equidistant and it is determined by sequence of Fibonacci divisors
\begin{equation}
E^{(k)}_n = \frac{\hbar \omega}{2} (F^{(k)}_n + F^{(k)}_{n+1}), \,\,\,n=0,1,2,...
\end{equation}
or
\begin{equation}
E^{(k)}_n = \frac{\hbar \omega}{2} (F_{n k+1} + F^{(k)}_n (F_{k-1} +1))\,.
\end{equation}
The energy levels satisfy the three terms recurrence relations
\begin{equation}
E^{(k)}_{n+1} = L_k E^{(k)}_n + (-1)^{k-1} E^{(k)}_{n-1}
\end{equation}
where the Lucas numbers are equal $L_k = \varphi^k + {\varphi'}^k$.
\subsection{Even $k$ Hierarchy}
For the even values of $k$ the spectrum of the system takes the form
\begin{equation}
E^{(k)}_n = \frac{\hbar \omega}{2} \frac{\sinh [(n+ \frac{1}{2}) k \ln \varphi]}{\sinh [\frac{1}{2} k \ln \varphi]}.
\end{equation}
In the limit $k \rightarrow 0$ this gives the energy levels of the linear harmonic oscillator
\begin{equation}
E^{(0)}_n = \hbar \omega \left(n + \frac{1}{2}\right).
\end{equation}
In another limit $k \rightarrow \infty$ the energy levels are growing exponentially as
\begin{equation}
E^{(k)}_n \approx \frac{\hbar \omega}{2} \varphi^{n k}\,.
\end{equation}
\subsection{Odd $k$ Hierarchy}
In case of odd $k$, for the even energy levels ($n = 2 s$) we have spectrum
\begin{equation}
E^{(k)}_n = \frac{\hbar \omega}{2} \sinh [(n+ \frac{1}{2}) k \ln \varphi]\,\frac{2\cosh [\frac{1}{2} k \ln \varphi]}{\cosh [ k \ln \varphi]}\,,
\end{equation}
while for the odd levels ($n = 2 s +1$) it is
\begin{equation}
E^{(k)}_n = \frac{\hbar \omega}{2} \cosh [(n+ \frac{1}{2}) k \ln \varphi]\,\frac{2\cosh [\frac{1}{2} k \ln \varphi]}{\cosh [ k \ln \varphi]}\,.
\end{equation}

In addition, for the odd $k$ we can define the hierarchy of Golden deformed fermionic oscillators with Hamiltonian
\begin{equation}
H_k =  \frac{\hbar \omega}{2} (b_k b^\dagger_k - b^\dagger_k b_k) = \frac{\hbar \omega}{2} (F^{(k)}_{N+I} - F^{(k)}_N ) \label{fermionicH}
\end{equation}
and the energy spectrum
\begin{equation}
E^{(k)}_n = \frac{\hbar \omega}{2} (F^{(k)}_{n+1} - F^{(k)}_n), \,\,\,n=0,1,2,...
\end{equation}
For $k=1$ case we have the Golden oscillator spectrum, determined by Fibonacci numbers
\begin{equation}
E^{(k)}_n = \frac{\hbar \omega}{2} (F_{n+1} - F_n) = \frac{\hbar \omega}{2} F_{n-1} , \,\,\,n=0,1,2,...
\end{equation}

\subsection{Hierarchy of Coherent States}

The hierarchy of Golden coherent states is defined by eigenstates of annihilation operator $b_k$,
\begin{equation}
b_k |\beta_k \rangle = \beta_k |\beta_k \rangle, \,\,\,\,\beta \in C.\label{GCS}
\end{equation}
By expanding in the deformed basis states (\ref{deformedFock}) we have explicit form for the normalized states
\begin{equation}
|0,\beta_k\rangle = \left(_{(k)}e^{|\beta_k|^2}_F\right)^{-1/2} |\beta_k \rangle    =  \left(_{(k)}e^{|\beta_k|^2}_F\right)^{-1/2}\sum^\infty_{n=0} \frac{\beta_k^n}{\sqrt{F^{(k)}_n !}} |n;k\rangle,
\end{equation}
with following inner product
\begin{equation}
\langle 0,\alpha_k |0, \beta_k \rangle = \frac{_{(k)} {e_F}^{\bar\alpha \beta}}{\sqrt{_{(k)} {e_F}^{|\alpha|^2} \,  _{(k)} {e_F}^{|\beta|^2}}}\,.
\end{equation}
Here, the hierarchy of the entire Golden exponential functions is defined as 
\begin{equation}
_{(k)}e^z_F = \sum^\infty_{n=0} \frac{z^n}{F^{(k)}_n!}.
\end{equation}
In the limit $k \rightarrow 0$,  Fibonacci divisors become the usual integer numbers $F^{(0)}_n = n$, the deformed exponential and the coherent states give the classical exponential function $_{(0)}e^z_F = e^z$ and the Glauber coherent states $|\beta_0\rangle = |\beta\rangle$, respectively, so that
$b |\beta\rangle = \beta|\beta\rangle$. 
For (not normalized) coherent states 
\begin{equation}
|\beta_k\rangle = \sum^\infty_{n=0} \frac{\beta_k^n}{\sqrt{F^{(k)}_n !}} |n;k\rangle = _{(k)}e^{\beta_k b_k^\dagger}|0;k\rangle,
\end{equation}
we can derive several relations:
\begin{eqnarray}
b_k |\frac{\beta_k}{\lambda}\rangle &=& \frac{\beta_k}{\lambda} |\frac{\beta_k}{\lambda}\rangle ,\\
   D_k |\beta_k\rangle &=& \sum^\infty_{n=1} \frac{F^{(k)}_n \beta_k^{n-1}}{\sqrt{F^{(k)}_n!}} |n;k\rangle \equiv |\beta'_k\rangle, \\
	D_k |\frac{\beta_k}{\lambda}\rangle &=& \sum^\infty_{n=1} \frac{F^{(k)}_n \beta_k^{n-1}}{\lambda^n\sqrt{F^{(k)}_n!}} |n;k\rangle \equiv |\frac{\beta_k'}{\lambda}\rangle, \\
	b_k |\frac{\beta'}{(-1)^k}\rangle &=& \frac{\beta_k}{\varphi^k}\, |\frac{\beta_k'}{(-1)^k}\rangle + (-1)^k |\frac{\beta_k}{{\varphi'}^k}\rangle \\
	&=&  
	\frac{\beta_k}{{\varphi'}^k}\, |\frac{\beta_k'}{(-1)^k}\rangle + (-1)^k |\frac{\beta_k}{{\varphi}^k}\rangle.
\end{eqnarray}
Here the derivative operator $D_k$, the main ingredient of quantum calculus,  is acting on function $f$ as
\begin{equation}
D_k f(x) = \frac{f(\varphi^k x) - f({\varphi'}^k x)}{(\varphi^k - {\varphi'}^k) x}.
\end{equation}
The inner products, where $\lambda, \mu$ are  real constants, are equal
\begin{equation}
\langle {\beta_k}| {\alpha_k}\rangle = \,_{(k)}e_F^{{\bar\beta_k\alpha_k}},\hskip0.5cm
\langle {\beta_k}| {\beta_k}\rangle = \,_{(k)}e_F^{{|\beta_k|^2}}, \hskip0.5cm
\langle \frac{\beta_k}{\lambda}| \frac{\beta_k}{\mu}\rangle = \,_{(k)}e_F^{\frac{|\beta_k|^2}{\lambda \mu}}, 
\end{equation}
\begin{eqnarray}
\langle \frac{\beta_k'}{\lambda}| \frac{\beta_k}{\mu}\rangle &=& {\frac{\beta_k}{\lambda \mu}}\, {_{(k)}e}_F^{\frac{|\beta_k|^2}{\lambda \mu}}, \hskip1cm
\langle \frac{\beta_k}{\mu}| \frac{\beta_k'}{\lambda}\rangle = {\frac{\bar\beta_k}{\lambda \mu}} \, {_{(k)}e}_F^{\frac{|\beta_k|^2}{\lambda \mu}}.
\end{eqnarray}
\begin{eqnarray}
\langle \frac{\beta_k'}{\lambda}| \frac{\beta_k'}{\mu}\rangle &=&
\frac{1}{\lambda\mu}\left(\frac{\varphi^k}{\lambda\mu}|\beta_k|^2 \,_{(k)}e_F^{\frac{|\beta_k|^2}{\lambda\mu}}   + \,_{(k)}e_F^{{\varphi'}^k\frac{|\beta_k|^2}{\lambda\mu}} \right)  \\
&=& \frac{1}{\lambda\mu}\left(\frac{{\varphi'}^k}{\lambda\mu}|\beta_k|^2 \,_{(k)}e_F^{\frac{|\beta_k|^2}{\lambda\mu}}   + \,_{(k)}e_F^{{\varphi}^k\frac{|\beta_k|^2}{\lambda\mu}} \right) .
\end{eqnarray}
It is noticed that in the last formula two representations of the inner product are equivalent.

 \section{Hierarchy of Supersymmetric Golden Oscillators}
 We introduce supersymmetric  operators 
\begin{equation}
Q_k = \left(\begin{array}{cc} 0 & 0 \\ b_k & 0 \end{array}\right), \hskip0.5cm Q^\dagger_k = \left(\begin{array}{cc} 0 & b^\dagger_k \\ 0 & 0 \end{array}\right),
\end{equation}
and corresponding Hamiltonian by following anti-commutator
\begin{equation}
H^s_k = \frac{\hbar \omega}{2}\{ Q_k, Q^\dagger_k \} = \frac{\hbar \omega}{2} (Q_k  Q^\dagger_k + Q^\dagger_k Q_k).  
\end{equation}
Then, in matrix form
\begin{equation}
H^s_k = \frac{\hbar \omega}{2}\left(\begin{array}{cc} b^\dagger_k b_k & 0 \\ 0 & b_k b^\dagger_k \end{array}\right) = \frac{\hbar \omega}{2}\left(\begin{array}{cc} F^{(k)}_N & 0 \\ 0 & F^{(k)}_{N+I} \end{array}\right). \label{H}
\end{equation}
The super-number operator is defined as
\begin{equation}
{\cal N} = \left(\begin{array}{cc} N & 0 \\ 0 & {N+I} \end{array}\right) = N_f \otimes I_b + I_f \otimes N,
\end{equation}
where $N = b^\dagger b$
and the supersymmetric Fibonacci divisors number operator is
\begin{equation}
{\cal F}^{(k)}_{\cal N} = \left(\begin{array}{cc} F^{(k)}_N & 0 \\ 0 & F^{(k)}_{N+I} \end{array}\right) = N_f \otimes (F^{(k)}_{N+I} - F^{(k)}_N) + I_f \otimes F^{(k)}_N.
\end{equation}
	It is noted that in this expression the structure of fermionic Golden oscillators (\ref{fermionicH}) appears in the first term of the sum.
	The operator ${\cal F}^{(k)}_{\cal N}$ is connected with supersymmetric number operator ${\cal N}$ by the Binet type formula
\begin{equation}
{{\cal F}^{(k)}_{\cal N}} = \frac{{\varphi}^{k {\cal N}} - {\varphi'}^{k {\cal N}}}{\varphi^k - {\varphi'}^k},
\end{equation}
	and the Hamiltonian is expressed in terms of ${\cal F}^{(k)}_{\cal N}$ as
	\begin{equation}
	H^s_k = \frac{\hbar \omega}{2} {\cal F}^{(k)}_{\cal N}.\label{HsuperFD}
	\end{equation}

	By taking partial trace in fermionic variables we get the Hamiltonian (\ref{goldenH}) for bosonic hierarchy of Golden oscillators
	\begin{equation}
Tr_f H^s_k =\frac{\hbar \omega}{2} (F^{(k)}_N + F^{(k)}_{N+I}) = H_k.
\end{equation}
This shows that the hierarchy of Golden oscillators $H_k$ in  (\ref{goldenH}) is reduction (by trace over fermionic degrees of freedom) of supersymmetric Hamiltonian $H^s_k$ (\ref{HsuperFD}).
The eigenstates of operators, ${{\cal F}^{(k)}_{\cal N}}$ and $H^s_k$ are the same, with eigenvalues given by Fibonacci divisors  number sequence $F^{(k)}_n$, so that
the spectrum of energy is 
\begin{equation}
E_n = \frac{\hbar \omega}{2} F^{(k)}_n.
\end{equation} 
The first few  sequences for hierarchy of energy levels  $E_n$ for $k=1,2,3,4,5$ and $n=1,2,3,4,5,...$ are given below
\begin{eqnarray}
 k &=& 1 ; E_n =\frac{\hbar \omega}{2} F_n = \frac{\hbar \omega}{2} , \frac{\hbar \omega}{2},2\frac{\hbar \omega}{2} ,3\frac{\hbar \omega}{2} ,5\frac{\hbar \omega}{2} ,...\\
 k& = &2 ; E_n =\frac{\hbar \omega}{2} F_{2n} =\frac{\hbar \omega}{2} ,3\frac{\hbar \omega}{2} ,8 \frac{\hbar \omega}{2},21\frac{\hbar \omega}{2} ,55\frac{\hbar \omega}{2} ,...\\
 k&= &3; E_n= \frac{\hbar \omega}{4}{F_{3n}}= \frac{\hbar \omega}{2},4 \frac{\hbar \omega}{2},17\frac{\hbar \omega}{2} ,72 \frac{\hbar \omega}{2}, 305\frac{\hbar \omega}{2} ,...\\
 k&=&4; E_n = \frac{\hbar \omega}{6}{F_{4n}}=\frac{\hbar \omega}{2} ,7\frac{\hbar \omega}{2} ,48\frac{\hbar \omega}{2} ,329\frac{\hbar \omega}{2} ,2255\frac{\hbar \omega}{2} ,...\\
 k&=&5; E_n = \frac{\hbar \omega}{10}{F_{5n}}=\frac{\hbar \omega}{2},11\frac{\hbar \omega}{2} ,122 \frac{\hbar \omega}{2},1353\frac{\hbar \omega}{2} ,15005\frac{\hbar \omega}{2} ,...
\end{eqnarray}

\subsection{Super Fibonacci Divisors Number States}
The eigenstates of operator ${{\cal F}^{(k)}_{\cal N}}$ are two types, 
\begin{equation}
{{\cal F}^{(k)}_{\cal N}} \left( \begin{array}{c} |n;k\rangle \\ 0 \end{array}   \right) = F^{(k)}_n \left( \begin{array}{c} |n;k\rangle \\ 0 \end{array}   \right),
\end{equation}
and 
\begin{equation}
{{\cal F}^{(k)}_{\cal N}} \left( \begin{array}{c} 0 \\ |n-1;k\rangle \end{array}   \right) = F^{(k)}_n \left( \begin{array}{c} 0 \\ |n-1;k\rangle \end{array}   \right),
\end{equation}
with number of fermions equal zero and one, correspondingly, and these states are separable. However, an arbitrary superposition of these states is also an eigenstate of ${{\cal F}^{(k)}_{\cal N}}$ and it could be entangled. 
By using normalization condition up to global phase, we define the supersymmetric Fibonacci divisors number state, as the double degenerate 
superposition of the states,  

\begin{equation}
|n;k, \theta, \phi \rangle = \cos \frac{\theta}{2} \left( \begin{array}{c} |n;k\rangle \\ 0 \end{array}   \right) + \sin \frac{\theta}{2} e^{i\phi}
\left( \begin{array}{c} 0 \\ |n-1;k\rangle \end{array}   \right). \label{numberstate}
\end{equation}
It is an eigenstate of ${{\cal F}^{(k)}_{\cal N}}$,
\begin{equation}
{{\cal F}^{(k)}_{\cal N}} |n;k, \theta, \phi \rangle = F^{(k)}_n |n;k, \theta, \phi \rangle,
\end{equation}
where $0\le \theta\le \pi, 0\le \phi\le 2\pi$.
The states can be represented by points on the unit sphere, which we call as the super-Bloch sphere. The north and the south poles of the sphere correspond to 
the zero and the one fermion states, which are the one qubit states according to the fermion number operator, and $n$ and $n-1$ number of bosons in deformed bosonic space
: $|0\rangle_f \otimes |n; k\rangle$ and $|1\rangle_f \otimes |n-1; k\rangle$. The total number of fermions and bosons, $n = n_f + n_b$, counts the number of super-particles,
which is the same for all states of the superposition on the sphere. Then, eigenvalues of the ${{\cal F}^{(k)}_{\cal N}}$ are $F^{(k)}_{n_f + n_b}$.

	The energy levels of the Hamiltonian (\ref{H}) in states $|n, \theta, \phi \rangle$ are given by Fibonacci divisors numbers
$E^{(k)}_n = \frac{\hbar \omega}{2}F^{(k)}_n$, where $n= n_f + n_b$ counts number of superparticles in the state. The energy levels satisfy the triple
addition rule 
\begin{equation}E^{(k)}_{n+1} = L_k E^{(k)}_n + (-1)^{k-1}E^{(k)}_{n-1}\end{equation}
and asymptotically for $n \rightarrow \infty$, they grow exponentially as $E_n \approx \varphi^{-k}e^{k n \ln \varphi}$.
The ratio of energy levels  $E_{n+1}/E_n = F_{n+1}/F_n \equiv \lambda_n$ satisfies equation
\begin{equation}
\lambda_{n+1} = L_k + \frac{(-1)^{k-1}}{\lambda_n}
\end{equation}
and in the limit $n \rightarrow \infty$ it is equal to the power of Golden Ratio, $\lambda_\infty = \varphi^k$, as a root  of quadratic equation 
$\lambda^2_\infty = L_k \lambda_\infty + (-1)^{k-1}$.
	\section{Hierarchy of Supersymmetric Golden Coherent States}
	Here we introduce the hierarchy of supersymmetric Golden coherent states as the eigenstates of super-annihilation operator. We do this by generalization of the operator, introduced in \cite{Zypman} (corresponding to $k=0$) and the one (when $k=1$) for the Golden oscillator in \cite{P6}, to the case of the Golden divisors quantum algebra \cite{P4}.
The set of Golden super-annihilation operators we define as
	\begin{equation}
	A_{\pm k} = \left(\begin{array}{cc} \varphi^k b_k & \pm 1 \\ 0 & {\varphi'}^k b_k \end{array}\right), \hskip0.5cm A^T_{\pm k} = \left(\begin{array}{cc} \varphi^k b_k & 0 \\ \pm 1 & {\varphi'}^k b_k \end{array}\right),
	\end{equation}
	and the corresponding eigenstates of these operators we call as the hierarchy of
	the Golden super-coherent states,
	\begin{equation}
	A_{\pm k} |\beta_k, L_{\mp}\rangle = \beta_k |\beta_k, L_{\mp}\rangle, \hskip0.5cm A^T_{\pm k} |\beta_k, B_{\mp}\rangle = \beta_k |\beta_k, B_{\mp}\rangle.
	\end{equation}
	By taking partial trace according to fermionic degrees of freedom, 
	 \begin{equation}Tr_f A_{\pm k} = (\varphi^k + {\varphi'}^k) b_k = L_k b_k\end{equation}
	we get annihilation operators $b_k$ multiple Lucas numbers $L_k$, and  the eigenvalue problem for super-coherent states reduces to the one for the hierarchy of Golden oscillator coherent states (\ref{GCS}), with eigenvalues $L_k \beta_k$.
	The $n$-th power of operator $A_{\pm k}$ is equal 
	\begin{equation}
	A^n_{\pm k} = \left(\begin{array}{cc} {\varphi}^{kn} b^n_k & \pm F^{(k)}_n b^{n-1}_k\\ 0 & {\varphi'}^{kn} b^n_k \end{array}\right),  
	\end{equation}
	or in another form
	\begin{equation}
	A^n_{\pm k} = F^{(k)}_n \left(\begin{array}{cc} {\varphi}^{k} b^n_k & \pm  b^{n-1}_k\\ 0 & {\varphi'}^{k} b^n_k \end{array}\right) + (-1)^{k+1} F^{(k)}_{n-1}
	\left(\begin{array}{cc} b^n_k & 0\\ 0 & b^n_k \end{array}\right),  
	\end{equation}
	and it can be used to generate higher order polynomial quantum algebras and corresponding supersymmetric multiphoton coherent states.

	The pair of normalized separable super-coherent Golden states is 
	\begin{eqnarray}
	|\beta_k, sep \uparrow\rangle = (_{(k)}e_F^{\frac{|\beta_k|^2}{\varphi^{2k}}})^{-1/2} \left( \begin{array}{c} |\frac{\beta_k}{\varphi^k}\rangle \\ 0 \end{array}   \right),
	\hskip1cm A_{\pm k} |\beta_k, sep \uparrow\rangle = \beta_k |\beta_k, sep \uparrow\rangle, \label{sep1}\\
	|\beta_k, sep \downarrow\rangle =  (_{(k)}e_F^{\frac{|\beta_k|^2}{{\varphi'}^{2k}}})^{-1/2} \left( \begin{array}{c} 0 \\  |\frac{\beta_k}{{\varphi'}^k}\rangle\end{array}   \right),
	\hskip1cm A^T_{\pm k} |\beta_k, sep \downarrow\rangle = \beta_k |\beta_k, sep \downarrow\rangle.\label{sep2}
	\end{eqnarray}

	For entangled super-coherent states we have the first pair of states
	\begin{equation}
	|\beta_k, L_\pm \rangle =  \frac{1}{\sqrt{\varphi^{2k}\, _{(k)} e_F^{\varphi^{2k} |\beta_k|^2} + \varphi^{k} |\beta_k|^2\, _{(k)} e_F^{|\beta_k|^2} + _{(k)} e_F^{{\varphi'}^{k} |\beta_k|^2}}}\left( \begin{array}{c} {(-1)}^k|\frac{\beta_k'}{(-1)^k}\rangle \\ \pm \varphi^k|\frac{\beta_k}{{\varphi'}^k} \rangle \end{array}   \right), \label{betaL}
	\end{equation}
	satisfying equation
	\begin{equation}
	A_{\pm k} |\beta_k, L_{\mp}\rangle = \beta_k |\beta_k, L_{\mp}\rangle,\label{Leigenvalue}
	\end{equation}
	and the second one
	\begin{equation}
	|\beta_k, B_\pm \rangle = \frac{1}{\sqrt{{\varphi'}^{2k}\, _{(k)}e_F^{{\varphi'}^{2k} |\beta_k|^2} + {\varphi'}^k |\beta_k|^2 \,_{(k)}e_F^{|\beta_k|^2} + _{(k)}e_F^{{\varphi^k} |\beta_k|^2} }}
	\left( \begin{array}{c} \pm {\varphi'}^k|\frac{\beta_k}{\varphi^k}\rangle \\  (-1)^k|\frac{\beta_k'}{(-1)^k} \rangle \end{array}   \right), \label{betaB}
	\end{equation}
	subject to eigenvalue problem
	\begin{equation}
	A^T_{\pm k} |\beta_k, B_{\mp}\rangle = \beta_k |\beta_k, B_{\mp}\rangle.
	\end{equation}
	\subsection{Symmetry Operator and Orthogonal Super-Coherent States}
	Let $S$ is an operator, commutative with super annihilation operator, so that $[A_{\pm k}, S] = A_{\pm k} S - S A_{\pm k} = 0$. Then, this operator can generate new eigenstate of the problem 
	(\ref{Leigenvalue}) from the given one. It follows from equality
	\begin{equation}
	A_{\pm k} |\beta_k, L_{\mp}\rangle = \beta_k |\beta_k, L_{\mp}\rangle \Rightarrow A_{\pm k} (S |\beta_k, L_{\mp}\rangle) = \beta_k (S |\beta_k, L_{\mp}\rangle)
	\end{equation}
	and operator $S$ is the symmetry operator. 
	For problem (\ref{Leigenvalue}) the symmetry operator is
	\begin{equation}
	S_{\mp k} = \left(\begin{array}{cc} {\varphi'}^k b_k & \mp 1 \\ 0 & {\varphi}^k b_k \end{array}\right),  
	\end{equation}
	so that $[A_{\pm k}, S_{\mp k}] =0$. Applying this operator to super-coherent states $|\beta_k, L_\mp\rangle$ in (\ref{betaL}), we get the new super-coherent states
	\begin{equation}
	S_{\mp k} |\beta_k, L_\mp\rangle = N_{\mp} \left( \begin{array}{c} {\varphi'}^{2k} \beta_k|\frac{\beta_k'}{(-1)^k}\rangle + L_k |\frac{\beta_k}{{\varphi'}^k} \rangle \\ 
	\mp (-1)^k \varphi^{3k} \beta_k|\frac{\beta_k}{{\varphi'}^k} \rangle \end{array}   \right).
	\end{equation}
	Successive applications of the symmetry operator  $S_{\mp k}$ to the state   $|\beta_k, L_\mp\rangle$   , produce the set of super-coherent states,  $S_{\mp k} |\beta_k, L_\mp\rangle$,  $S^2_{\mp k} |\beta_k, L_\mp\rangle$, ...., $S^n_{\mp k} |\beta_k, L_\mp\rangle$.
	For the powers of this operator we have compact expression
	\begin{equation}
	S^n_{\mp k} = \left(\begin{array}{cc} {\varphi'}^{kn} b^n_k & \mp F^{(k)}_n b^{n-1}_k\\ 0 & {\varphi}^{kn} b^n_k \end{array}\right),  
	\end{equation}
	in terms of Fibonacci divisors $F^{(k)}_n$. It can be proved by induction. Then, by using formula
	\begin{equation}
	\varphi^{kn} = \varphi^k F^{(k)}_n + (-1)^{k+1} F^{k}_{n-1}
	\end{equation} 
	for the $n$-th power we obtain another representation 
	\begin{equation}
	S^n_{\mp k} = F^{(k)}_n \left(\begin{array}{cc} {\varphi'}^{k} b^n_k & \mp  b^{n-1}_k\\ 0 & {\varphi}^{k} b^n_k \end{array}\right) + (-1)^{k+1} F^{(k)}_{n-1}
	\left(\begin{array}{cc} b^n_k & 0\\ 0 & b^n_k \end{array}\right).  
	\end{equation}
	\subsection{Supersymmetric Golden Reference States}
In the limit $\beta_k \rightarrow 0$, we have the limiting state $|\frac{\beta_k'}{(-1)^k} \rangle \rightarrow (-1)^k|1;k\rangle $, giving 
	four reference states
	\begin{eqnarray}
	|\beta_k, L_\pm \rangle \rightarrow  | L_\pm\rangle  &=&\frac{|0\rangle_f |1;k\rangle \pm \varphi^k |1\rangle_f |0;k\rangle}{\sqrt{1 + \varphi^{2k}}}, \label{L}\\
	|\beta_k, B_\pm \rangle \rightarrow  |B_\pm \rangle &=&  \frac{\pm {\varphi'}^k|0\rangle_f |0;k\rangle + |1\rangle_f |1;k\rangle}{\sqrt{1 + {\varphi'}^{2k}}}, \label{B}
	\end{eqnarray}
	from coherent states (\ref{betaL}), (\ref{betaB}).
	These states are annihilated by corresponding super-annihilation operators
	\begin{equation}
	   A_{\pm k} |L_{\mp}\rangle = 0,\hskip1cm       A^T_{\pm k} |B_{\mp}\rangle = 0,
	\end{equation}
	and it is natural to call them as the hierarchy of Golden reference states.
	The first pair of reference states (\ref{L}) represents the super-number states (\ref{numberstate}) with one super-particle
	$
	{\cal F_N} | L_\pm\rangle = 1 | L_\pm\rangle
	$, located on the super-Bloch sphere at point with coordinates $tan \frac{\theta}{2} = \varphi$, $\phi = 0, \pi $.
	Stereographic projection of the sphere to the complex plane $\xi$ is implemented by transformation $\xi = \tan \frac{\theta}{2} e^{i\phi}$, so that
	states $| L_\pm\rangle_F$ correspond to real numbers $\xi = \varphi^k$ and $\xi =-\varphi^k$, representing powers of the Golden Ratio.

	The separable coherent  states (\ref{sep1}) and (\ref{sep2}) in the limit $\beta_k \rightarrow 0$, also determine the reference states, since
\begin{equation}
	   A_{\pm k} |0, sep \uparrow\rangle =0,\hskip1cm       A^T_{\pm k} |0, sep \downarrow\rangle = 0.
	\end{equation}
This implies that a linear combination of the states is also the reference state
\begin{equation}
	   A_{\pm k} (c_0 |0, sep \uparrow\rangle + c_1  |L_{\mp}\rangle  )=0     ,\hskip1cm   
				A^T_{\pm k} (d_0 |0, sep \downarrow\rangle + d_1 |B_{\mp}\rangle) = 0.\nonumber
	\end{equation}
In a similar way, by taking linear combination of separable and entangled super-coherent states, we get entangled super-coherent states as
\begin{equation}
 A_{\pm k} (c_0 |\alpha, sep \uparrow\rangle + c_1  |\alpha, L_{\mp}\rangle )= \alpha(c_0 |\alpha, sep \uparrow\rangle + c_1  |\alpha, L_{\mp}\rangle  ),
\end{equation}
\begin{equation}
A^T_{\pm k} (d_0 |\alpha, sep \downarrow\rangle + d_1 |\alpha, B_{\mp}\rangle) = \alpha (d_0 |\alpha, sep \downarrow\rangle + d_1 |\alpha, B_{\mp}\rangle).
\end{equation}
	\section{Entanglement and Entropy of the States}
		The product state $|\Psi\rangle_f \otimes |\Phi\rangle_b$ from fermion-boson Hilbert space ${\cal H}_f \otimes {\cal H}_b$ is the fermion-boson separable state. 
	If a state is not separable, then it is entangled. 
	To calculate fermion-boson entanglement of supersymmetric states we will use expression for concurrence 
	in the form of Gram determinant of inner products \cite{P5} in Fock space. Since deformed and non-deformed number states in Fock space coincide, $|n\rangle = |n;k\rangle$,
	we can adopt an approach in \cite{P5}  to the states from $H_f \otimes H_b$.

	The fermionic-bosonic  basis states are formed
by tensor product of fermionic (qubit) states with Fock states, $|0\rangle_f \otimes |k\rangle$, and $|1\rangle_f \otimes |k \rangle$, $k=0, 1, 2,...$ and 
for generic state
\begin{equation}
|\Psi \rangle = \sum^\infty_{k=0} c_{0 k} |0\rangle_f \otimes |k\rangle +  \sum^\infty_{k=0} c_{1 k} |1\rangle_f \otimes |k\rangle, \label{BFstate}
\end{equation}
from ${\cal H}_f \otimes {\cal H}_b$ Hilbert space we have  two representations. The first one 
\begin{eqnarray}
|\Psi\rangle = |0\rangle_f \otimes |\psi_0\rangle + |1\rangle_f \otimes |\psi_1\rangle = \left(
          \begin{array}{c}
            |\psi_0\rangle  \\
            |\psi_1 \rangle        \\
          \end{array}
        \right), \label{firstrepresentation}
\end{eqnarray}
is determined by
two bosonic states
\begin{eqnarray}
|\psi_0 \rangle = \sum^{\infty}_{k=0} c_{0 k} |k \rangle, \,\,\,\,\,|\psi_1 \rangle = \sum^{\infty}_{k=0} c_{1 k} |k \rangle. \label{bosonicstates}
\end{eqnarray}
as vectors  in the Fock space.
The second one
\begin{equation}
|\Psi \rangle = |\varphi_0\rangle_f \otimes |0\rangle + |\varphi_1\rangle_f \otimes |1\rangle + ... + |\varphi_{n}\rangle_f \otimes |0\rangle + ... = \sum^{\infty}_{n=0} |\varphi_n\rangle_f \otimes |n\rangle ,
\end{equation}
is characterized by infinite set of qubits $|\varphi_n \rangle_f$, $n=0,1,2,...$,
defined as
\begin{eqnarray}
|\varphi_{n}\rangle_f = \left(
          \begin{array}{c}
            c_{0 n}  \\
            c_{1 n}        \\
          \end{array}
        \right) = c_{0 n} |0\rangle_f + c_{1 n} |1\rangle_f .\label{infinitequbits}
\end{eqnarray}
	
For normalized state (\ref{firstrepresentation})
the density matrix $\rho = | \Psi \rangle \langle \Phi |$,
 due to normalization condition, gives
\begin{equation}
tr \rho = \langle \psi_0 | \psi_0 \rangle + \langle \psi_1 | \psi_1 \rangle = 1. \label{tracerho}
\end{equation}
 By taking partial trace, for the reduced bosonic density matrix
\begin{equation}
\rho_b = tr_f \,\rho = | \psi_0 \rangle \langle \psi_0| + | \psi_1 \rangle \langle \psi_1|
\end{equation}
 we find
\begin{equation}
tr \rho^2_b = \langle \psi_0 | \psi_0 \rangle^2 + \langle \psi_1 | \psi_1 \rangle^2 + 2 |\langle \psi_0 | \psi_1 \rangle|^2,
\end{equation}
and for the fermionic one
\begin{equation}
\rho_f = tr_b \,\rho = \sum^\infty_{n=0} |\varphi_n \rangle \langle \varphi_n | \label{reduceddensitymatrixf}
\end{equation}
it is
\begin{equation}
tr \rho^2_f = \sum^\infty_{n=0} \sum^\infty_{m=0}    |\langle \varphi_n | \varphi_m \rangle|^2.
\end{equation}
As easy to check by direct computation,  it coincides with the bosonic one,  so that $tr \rho^2_b = tr \rho^2_f$.
The first expression we rewrite in the form
\begin{equation}
tr \rho^2_b = (\langle \psi_0 | \psi_0 \rangle + \langle \psi_1 | \psi_1 \rangle)^2 - 2 (\langle \psi_0 | \psi_0 \rangle \langle \psi_1 | \psi_1 \rangle - \langle \psi_0 | \psi_1 \rangle \langle \psi_1 | \psi_0 \rangle)
\end{equation}
and by subtracting it from the squared equation (\ref{tracerho}), we get
\begin{equation}
1 - tr \rho^2_b = 2 \left|
 \begin{array}{cc}
  \langle\psi_0 | \psi_0 \rangle &        \langle\psi_0 | \psi_1 \rangle       \\ \\
     \langle\psi_1 | \psi_0 \rangle &  \langle\psi_1 | \psi_1 \rangle  \\
	\end{array} \right|.     \label{difference}
\end{equation}
Deviation of this trace from unity gives a simplest characteristics of the level of entanglement,
known as the linear entropy. For a pure fermion-boson state, the concurrence $C$ is defined by relation
\begin{equation}
    \frac{C^2}{2} \equiv  1 - tr \rho^2_b   \hskip0.5cm     \Rightarrow       \hskip0.5cm       tr \rho_b^2 + \frac{1}{2} C^2 =1\,, \label{concurrence}
\end{equation}	
	or
\begin{equation}
C = \sqrt{2} \sqrt{1- tr \rho_b^2}.\label{C1}
\end{equation}
From (\ref{difference}), (\ref{concurrence}) we find the concurrence square as determinant of the Hermitian inner product metric
$g_{ij} = \langle\psi_i | \psi_j \rangle$, (the Gram determinant), of two vectors $(i,j = 0,1)$ in Fock space,
\begin{equation}
C^2 = 4\left|
 \begin{array}{cc}
  \langle\psi_0 | \psi_0 \rangle &        \langle\psi_0 | \psi_1 \rangle       \\ \\
     \langle\psi_1 | \psi_0 \rangle &  \langle\psi_1 | \psi_1 \rangle  \\
	\end{array} \right|,\label{C2b}
\end{equation}
and for the generic quantum state (\ref{BFstate}),
\begin{equation}
C = 2 \,\sqrt{det \left(
 \begin{array}{cc}
  \langle\psi_0 | \psi_0 \rangle &        \langle\psi_0 | \psi_1 \rangle       \\ \\
     \langle\psi_1 | \psi_0 \rangle &  \langle\psi_1 | \psi_1 \rangle  \\
	\end{array} \right) }. \label{genericconcurrence}
\end{equation}
Similar calculations for reduced density matrix $\rho_f$ give relation
\begin{equation}
tr \rho^2_f = 1 - \sum^\infty_{n=0} \sum^\infty_{m=0} \left|
 \begin{array}{cc}
  \langle\varphi_n | \varphi_n \rangle &        \langle\varphi_n | \varphi_m \rangle       \\ \\
     \langle\varphi_m | \varphi_n \rangle &  \langle\varphi_m | \varphi_m \rangle  \\
	\end{array} \right|,
\end{equation}
and another representation of the concurrence 
\begin{equation}
C^2 =2 \sum^\infty_{n=0} \sum^\infty_{m=0} \left|
 \begin{array}{cc}
  \langle\varphi_n | \varphi_n \rangle &        \langle\varphi_n | \varphi_m \rangle       \\ \\
     \langle\varphi_m | \varphi_n \rangle &  \langle\varphi_m | \varphi_m \rangle  \\
	\end{array} \right|.\label{C2}
\end{equation}
Comparing two representations of concurrence (\ref{C2b}) and (\ref{C2}),  we get identity
\begin{equation}
\left|
 \begin{array}{cc}
  \langle\psi_0 | \psi_0 \rangle &        \langle\psi_0 | \psi_1 \rangle       \\ \\
     \langle\psi_1 | \psi_0 \rangle &  \langle\psi_1 | \psi_1 \rangle  \\
	\end{array} \right| = \sum^\infty_{0=n < m}
	\left|
 \begin{array}{cc}
  \langle\varphi_n | \varphi_n \rangle &        \langle\varphi_n | \varphi_m \rangle       \\ \\
     \langle\varphi_m | \varphi_n \rangle &  \langle\varphi_m | \varphi_m \rangle  \\
	\end{array} \right|.
\end{equation}
By using explicit form of the one qubit states (\ref{infinitequbits}), the concurrence square can be rewritten as an infinite sum of modulus squares of all $2\times 2$ minors of the coefficient matrix $c_{n m}$,
\begin{equation}
C^2 =4 \sum^\infty_{0=n < m} \left| \left|
 \begin{array}{cc}
  c_{0 n} &   c_{0 m}      \\ \\
     c_{1 n} &  c_{1 m} \\
	\end{array} \right|\right|^2.\label{concurrenceC}
\end{equation}
Then,
for generic normalized fermion-boson state (\ref{BFstate}) from Hilbert space ${\cal H}_f \otimes {\cal H}_b$, the concurrence is equal
\begin{equation}
C= 2 \,\sqrt{det \left(
 \begin{array}{cc}
  \langle\psi_0 | \psi_0 \rangle &        \langle\psi_0 | \psi_1 \rangle       \\ \\
     \langle\psi_1 | \psi_0 \rangle &  \langle\psi_1 | \psi_1 \rangle  \\
	\end{array} \right)}
=2 \,\sqrt{ \sum^\infty_{0=n < m} \left| \left|
 \begin{array}{cc}
  c_{0 n} &   c_{0 m}      \\ \\
     c_{1 n} &  c_{1 m} \\
	\end{array} \right|\right|^2 }. \label{Cformula}
\end{equation}
As a result,
the determinant of $2\times 2$ inner product metric in Fock space can be represented by an infinite sum of modulus squares of minors of the infinite matrix from
coefficients $c_{n m}$ of the state (\ref{BFstate}),
\begin{eqnarray}
det \left(
 \begin{array}{cc}
  \langle\psi_0 | \psi_0 \rangle &        \langle\psi_0 | \psi_1 \rangle       \\ \\
     \langle\psi_1 | \psi_0 \rangle &  \langle\psi_1 | \psi_1 \rangle  \\
	\end{array} \right)  =  \sum^\infty_{0=n < m} \left| \left|
 \begin{array}{cc}
  c_{0 n} &   c_{0 m}      \\ \\
     c_{1 n} &  c_{1 m} \\
	\end{array} \right|\right|^2.
\end{eqnarray}

\subsection{Entanglement of Super-Number States}
Now we can calculate entanglement for super-number states.
In arbitrary state (\ref{numberstate}), except north and south poles,  the fermionic and bosonic degrees of freedom are entangled. To calculate the level of entanglement 
we start from density operator $\rho_n = | n;k, \theta, \phi\rangle_F {_F}\langle n;k, \theta, \phi|$ and by getting partial trace in fermionic variables we obtain the reduced density matrix 
\begin{equation}
\rho_b = tr_f \rho_n  = \sin^2 \frac{\theta}{2} |n-1;k\rangle_F {_F}\langle n-1;k| + \cos^2 \frac{\theta}{2} |n;k\rangle_F {_F}\langle n;k|.
\end{equation}
  Taking trace $tr \rho^2_b = 1 - \frac{1}{2} \sin^2 \theta$ and using definition of the concurrence (\ref{C1}), we
	find 
	\begin{equation}
	C = \sin \theta\,.
	\end{equation}
	From this formula, the maximally fermion-boson entangled states correspond to $\theta = \pi/2$ and they are located on the big circle of equator. 

	\subsection{Entanglement of Reference States}
	For states (\ref{L}), the bosonic representation states are
	\begin{equation}
	|\psi_0\rangle= \frac{1}{\sqrt{1 + \varphi^{2k}}} |1;k\rangle, \hskip1cm |\psi_1\rangle = \pm\frac{\varphi^k}{\sqrt{1 + \varphi^{2k}}} |0;k\rangle,\label{bosstates}
	\end{equation}
	while for states (\ref{B}), we have
	\begin{equation}
	|\psi_0\rangle = \pm\frac{{\varphi'}^k}{\sqrt{1 + {\varphi'}^{2k}}} |0;k\rangle, \hskip1cm |\psi_1\rangle = \frac{1}{\sqrt{1 + {\varphi'}^2}} |1;k\rangle.
	\end{equation}
	By calculating the inner products in the Gram determinant and using the Golden Ratio identity $\varphi^2 = 1 + \varphi$, and $\varphi' = -1/\varphi$, we get the following result.
	
	For the reference states (\ref{L}) and (\ref{B}),  the concurrence takes the same value, determined by power of the Golden Ratio, 
	\begin{equation}
	C = \frac{2\, \varphi^k}{1 + \varphi^{2k}} = 2\frac{\varphi F_k + F_{k-1}}{\varphi F_{2k} + F_{2k-1} +1}.\label{C}
	\end{equation}
	For $k=0$ the reference states are maximally entangled Bell states with $C=1$. For $k=1$ they are highly entangled with $C \approx 0.894$.
	When $k \rightarrow \infty$, the states become separable.

	The von Neumann entropy, as the measure of entanglement for supersymmetric states is determined by concurrence completely,
	\begin{eqnarray}
	E = - \frac{1 + \sqrt{1-C^2}}{2} \log_2 \frac{1 + \sqrt{1-C^2}}{2} - \frac{1 - \sqrt{1-C^2}}{2} \log_2 \frac{1 - \sqrt{1-C^2}}{2}.
	\end{eqnarray}
	By substituting $C$ from (\ref{C}) to this formula we get the value of entanglement for the hiearchy of Golden reference states
	(\ref{L}) and (\ref{B}). It is given by value of the von Neumann entropy,
	expressed by powers of Golden Ratio only,
	\begin{equation}
	E = \log_2 (\varphi^{2k} +1) - 2 \frac{\varphi^{2k}}{\varphi^{2k} +1}\log_2 \varphi^{k} = \log_2 \frac{\varphi^{2k} +1}{(\varphi^{2k})^{\frac{\varphi^{2k}}{\varphi^{2k}+1}}}.
	\end{equation}
	\subsection{Frobenius Spherical Shell for Entangled States}
	Here we present geometrical interpretation of separable and entangled states as points in spherical shell. We start from 
	qubit$\otimes$n-qudit state from Hilbert space ${\cal H}_f \otimes {\cal H}_n$ with density matrix $\rho$. By taking 
	partial trace in fermionic degrees of freedom we get the reduced density matrix
	$
	\rho_n = tr_f \rho
	$,
	so that
	$ tr \rho_n = 1$. The level of mixture in this state is determined by the Frobenius norm
	\begin{equation}
	tr \rho^2_n = tr \rho^\dagger_n \rho_n = \sum^{n-1}_{i,j=0}|(\rho_n)_{ij}|^2 \equiv || \rho_n ||_F^2
	\end{equation}
	for Hermitian matrix $\rho_n = \rho^\dagger_n$. Then the linear entropy $S_L$ and the concurrence $C$ can be expressed by this norm
	\begin{equation}
	S_L = 1 - tr \rho^2_n = 1 - ||\rho_n ||^2 = \frac{1}{2} C^2,
	\end{equation}
	so that
	\begin{equation}
	C^2 = 2 (1 - ||\rho_n ||^2), \label{Fconcurrence}
	\end{equation}
	For separable states $tr \rho^2_n =1$, which implies $C = 0$ and 
	\begin{equation}
	||\rho_n ||^2 =1. \label{Fsphere}
	\end{equation}
	This equation represents the unit length sphere, which we call as the Frobenius sphere for reduced density matrix, so that every state on this sphere 
	corresponds to separable state from  ${\cal H}_f \otimes {\cal H}_n$. If $tr \rho^2_n \neq 1$, the concurrence $C \neq 0$ and the original states are
	entangled. The level of entanglement can be characterized by concurrence. From (\ref{Fconcurrence}) and $tr \rho_n =1$,
	by using Frobenius norm
	\begin{equation}
	|| \rho_n ||^2 = \sum^{n-1}_{i=0} (\rho_n)^2_{ii} + 2 \sum^{n-1}_{0 = i <j} |(\rho_n)_{ij}|^2
	\end{equation}
	we have
	\begin{eqnarray}
	C^2 &=& 2 (1 - ||\rho_n ||^2)\\ &=& 2 (\sum^{n-1}_{i=0} (\rho_n)_{ii}   \sum^{n-1}_{j=0} (\rho_n)_{jj} -  
	\sum^{n-1}_{i=0} (\rho_n)^2_{ii} - 2 \sum^{n-1}_{0 = i <j} |(\rho_n)_{ij}|^2) \\
	&=& 4 \sum^{n-1}_{0 = i <j} [(\rho_n)_{ii} (\rho_n)_{jj} - (\rho_n)_{ij} (\bar\rho_n)_{ij} ]
	\end{eqnarray}
	or 
	\begin{equation}
C^2 = 4 \sum^{n-1}_{0 = i <j}     \left|
 \begin{array}{cc}
  (\rho_n)_{ii} &   (\rho_n)_{ij}        \\ \\
      (\rho_n)_{ji} &  (\rho_n)_{jj} \\
	\end{array} \right|.\label{CF}
\end{equation}
In this formula the sum is going according to all $2\times 2$ minors of the reduced density matrix.
To find maximal value of concurrence for entangled states, we diagonalize Hermitian matrix $\rho_n$ by
a unitary matrix $U$, so that
\begin{equation}
\rho_n = U\, \Lambda\, U^\dagger
 \end{equation}
	and $\Lambda = diag(\lambda_0, \lambda_1,...,\lambda_{n-1})$, with real eigenvalues. For unit trace condition
	we have 
	\begin{equation}
	\sum^{n-1}_{i =0} \lambda_i = 1
	\end{equation}
	and for Frobenius norm
	\begin{equation}
	||\Lambda ||_F^2 = \sum^{n-1}_{i=0} \lambda^2_{i}. 
	\end{equation}
	This gives the concurrence square as
	\begin{equation}
	C^2 = 2 (1 -  \sum^{n-1}_{i=0} \lambda^2_{i} ) = 4 \sum^{n-1}_{0 = i <j} \lambda_i \lambda_j.
	\end{equation}
	To evaluate the maximal value of the concurrence we consider function
	\begin{eqnarray}
	F(\lambda_0,...,\lambda_{n-1},\lambda) &=& C^2 - \lambda ((tr \Lambda)^2 -1),\\
	&=&  2 (1 -  \sum^{n-1}_{i=0} \lambda^2_{i} ) - \lambda ((\sum^{n-1}_{i=0} \lambda_i)^2 -1)
	\end{eqnarray}
	For critical points we have the system ($i = 0, 1,...,n-1$)
	\begin{equation}
	\frac{\partial F}{\partial \lambda_i} = - 4 \lambda_i - 2 \lambda \sum^{n-1}_{i=0} \lambda_i =0,
	\end{equation}
	with solution
	\begin{equation}
	\lambda_0 = \lambda_1 =...= \lambda_{n-1} = \frac{1}{n}
	\end{equation}
	Then, for maximally entangled states 
	\begin{equation}
	||\Lambda ||^2_F = \frac{1}{n}
	\end{equation}
	and the maximal value of concurrence is
	\begin{equation}
	C^2_{max} = 2 \left(1 - \frac{1}{n}\right) = 2\, \frac{n-1}{n}.\label{Cmax}
	\end{equation}
	In first few particular cases we have:1) $n=2$, the qubit-qubit state, $C_{max} = 1$; 2) $n=3$, the qubit-qutrit state, $C_{\max} = 2/\sqrt{3}$; 3) $n=4$, 
	the qubit-ququad state, $C_{max} = \sqrt{3}/\sqrt{2}$.
	Then, for fermion-boson states we have to take limit $n \rightarrow \infty$, giving the maximall concurrence 
	\begin{equation}
	C_{\max} = \sqrt{2}.
	\end{equation}
	The above calculations allow us to find geometrical image of entangled states. In fact, the maximal concurrence (\ref{Cmax})
	implies the Frobenius sphere 
	\begin{equation}
	||\Lambda ||^2_F = \frac{1}{n}
	\end{equation}
	with radius $r = 1/\sqrt{n}$. As a result, we have geometrical classification of states. The separable qubit-n-qudit states are determined by 
	Frobenius sphere with radius one, $|| \rho_n ||^2 =1$. The entangled states belong to the spherical shell with radius $1/\sqrt{n} \leq r \le 1$.
	For fermion-boson case, when $n \rightarrow \infty$, the radius of sphere $r = 1/\sqrt{n}$, for maximally entangled  states, vanishes $r =0$
	and we have only one state at the origin. This is why, entangled fermion-boson states belong to the Frobenius ball $|| \rho_n ||^2 < 1$
	with maximal value  $\sqrt{2}$ for concurrence, at the center of the ball.
	\subsection{Geometric Interpretation of Concurrence}
	The concurrence formula for real states admits simple geometrical interpretation. Let us consider two states (\ref{bosonicstates}),
	\begin{eqnarray}
|\vec{a} \rangle = \sum^{\infty}_{k=0} a_{k} |k \rangle, \,\,\,\,\,|\vec{b} \rangle = \sum^{\infty}_{k=0} b_{k} |k \rangle, \label{realbosonicstates}
\end{eqnarray}
	from the Fock space, with real coefficients $a_k, b_k$, $k=0,1,2,...$. With these states we can associate two real vectors $\vec{a} = (a_0, a_1, a_2, ...)$ and 
	$\vec{b} = (b_0, b_1, b_2,...)$ from Euclidean space $E_\infty = \lim_{n \rightarrow \infty} E_n$. These vectors determine the parallelogram with area
	$A = |{\vec{a}}| |{\vec{b}}| \sin \alpha = \sqrt{\vec{a}^2 \vec{b}^2 - (\vec{a}\cdot \vec{b})^2}$, which is the square root of Gram determinant for states (\ref{realbosonicstates}).
	The determinant can be expanded as
	\begin{eqnarray}
	det \left(
 \begin{array}{cc}
  (\vec{a} \cdot \vec{a}) &      (\vec{a} \cdot \vec{b})       \\ \\
    (\vec{a} \cdot \vec{b})  &  (\vec{b} \cdot \vec{b})  \\
	\end{array} \right) = \sum_{n=0}^{\infty} a_n^2 \sum_{m=0}^{\infty} b_m^2 - \sum_{n=0}^{\infty} a_n b_n \sum_{m=0}^{\infty} a_m b_m \\
	= \sum_{n=0}^\infty \sum_{m=0}^\infty a_n b_m (a_n b_m - a_m b_n) = \frac{1}{2} \sum_{n=0}^\infty \sum_{m=0}^\infty (a_n b_m - a_m b_n)^2
	\end{eqnarray}
	or
	\begin{eqnarray}
	A^2 = det \left(
 \begin{array}{cc}
  (\vec{a} \cdot \vec{a}) &      (\vec{a} \cdot \vec{b})       \\ \\
    (\vec{a} \cdot \vec{b})  &  (\vec{b} \cdot \vec{b})  \\
	\end{array} \right) =  \sum_{0 = n < m}^\infty  \left|
 \begin{array}{cc}
  a_n &    a_m       \\ \\
    b_n  &  b_m  \\
	\end{array} \right|^2
	\end{eqnarray}
	This relation plays role of Pythagoras theorem for areas in $E_\infty$,
	\begin{equation}
	A^2 = \sum_{0 = n < m}^\infty A^2_{nm},
	\end{equation}
	where every term in summation is square of the area of the parallelogram $\{\vec{a},\vec{b}\}$ projection
	to the plane with axes $X_n$ and $X_m$,
	\begin{equation}
	A_{nm} = \left|
 \begin{array}{cc}
  a_n &    a_m       \\ \\
    b_n  &  b_m  \\
	\end{array} \right|.
	\end{equation}
	Comparing this formula with (\ref{concurrenceC}) we observe that $C^2 = 4 A^2$, so that
	\begin{equation}
	C = 2 A.
	\end{equation}
	The last expression shows that for the real quantum states the concurrence has simple geometrical meaning of the double area of parallelogram,
	determined by states $|\vec{a}\rangle$ and $|\vec{b}\rangle$ in $E_\infty$, playing role of the Hilbert space.
	The states (\ref{L}) and (\ref{B}) are real states, this is why we can apply this geometric interpretation to these states. For states (\ref{L})
	we have two real vectors (\ref{bosstates}) with concurrence (\ref{C}). It implies that these states determine the parallelogram with two orthogonal vectors
	\begin{equation}
	\vec{a} = (0, \frac{1}{\sqrt{1 + \varphi^{2k}}}, 0,...), \,\,\,\,\vec{b} = (\frac{\pm \varphi^k}{\sqrt{1 + \varphi^{2k}}}, 0, 0,...),
	\end{equation}
	and the area
	\begin{equation}
	A = \frac{\varphi^k}{1 + \varphi^{2k}}.
	\end{equation}
	The ratio of sights of the rectangle is
	\begin{equation}
	\frac{|\vec{b}|}{|\vec{a}|} = \varphi^k = \varphi F_k + F_{k-1}.
	\end{equation}
	For $k=0$, we have the square with unit diagonal, corresponding to maximally entangled state and $C=1$. For $k=1$, the rectangle becomes the Golden rectangle, with Golden ratio $\varphi$ of the sights.
		For arbitrary $k$ this ratio is growing by formula given above as Fibonacci numbers.
	\subsection{Entanglement of Supersymmetric Hierarchy of Golden Coherent States}
	The concurrence for hierarchy of supersymmetric Golden coherent states $|\beta_k, L_\pm \rangle$ is equal
	\begin{equation}
	C = 2 \varphi^k  \frac{\sqrt{e_F^{\varphi^k |\beta_k|^2} e_F^{{\varphi}^{2k} |\beta_k|^2} + {\varphi'}^k |\beta_k|^2 e_F^{ |\beta_k|^2} e_F^{{\varphi}^{2k} |\beta_k|^2} - 
	\varphi^{2k}|\beta_k|^2(e_F^{\varphi^k |\beta_k|^2})^2  }}{\varphi^{2k} e_F^{\varphi^{2k} |\beta_k|^2} + \varphi^k |\beta_k|^2 e_F^{|\beta_k|^2} + e_F^{{\varphi'}^k |\beta_k|^2}}.
	\end{equation}
	and for the states $|\alpha, B_\pm \rangle_F$ it is
	 \begin{equation}
	C = 2 |\varphi'|^k  \frac{\sqrt{e_F^{{\varphi'}^k |\beta_k|^2} e_F^{{\varphi'}^{2k} |\beta_k|^2} + {\varphi}^k |\beta_k|^2 e_F^{ |\beta_k|^2} e_F^{{\varphi'}^{2k} |\beta_k|^2} - 
	{\varphi'}^{2k}|\beta_k|^2(e_F^{{\varphi'}^k |\beta_k|^2})^2  }}{{\varphi'}^{2k} e_F^{{\varphi'}^{2k} |\beta_k|^2} + {\varphi'}^k |\beta_k|^2 e_F^{|\beta_k|^2} + e_F^{{\varphi}^k |\beta_k|^2}}.
	\end{equation}
	In these formulas we skip index $k$ for exponential functions.
	The proof is straightforward calculation of inner products for associated Golden boson coherent states and the corresponding Gram determinants.
	In the limit $\beta_k \rightarrow 0$ the concurrences take form of the one for the reference states
	\begin{equation}
	C = \frac{2\varphi^k}{1 + {\varphi}^{2k}}.
	\end{equation}
	%
%
% ---- Bibliography ----
%

\end{document}